\documentclass[showpacs,superscriptaddress, preprintnumbers,amsmath,amssymb,twocolumn]{revtex4}

\usepackage{graphicx}
\usepackage{dcolumn}
\usepackage{bm}
\include{epsf}

\begin{document}

\preprint{ preprint for PRE}

\title{Generalized Hurst exponent and multifractal function of original and translated texts mapped into frequency and length time series.
 }
\author{ M.
Ausloos }  \email{marcel.ausloos@ulg.ac.be} 
\address{483/0021 rue de la belle jardini\`ere, B-4031 Li\`ege Angleur, Belgium, Europe
\\{\it $previously$ at} GRAPES@SUPRATECS, Universit\'e
de Li\`ege, \\ Sart-Tilman, B-4000 Li\`ege, Euroland} %

  \date{\today}
 
\begin{abstract}
A nonlinear dynamics approach can be used in order to quantify complexity in written texts. As a first step,  a one-dimensional system is examined :  two written texts by one author (Lewis Carroll) are considered, together with one translation, into an artificial language, i.e. Esperanto are mapped into time series. Their corresponding shuffled  versions are used for obtaining a ''base line''.  Two different  one-dimensional time series are used here:  (i) one   based on  word lengths (LTS),  (ii) the other on word frequencies (FTS). It is shown that the generalized Hurst exponent $h(q)$  and the  derived  $f(\alpha)$ curves  of
the original and translated texts  show marked differences. The original "texts"  are far from giving a parabolic $f(\alpha)$ function, - in contrast to  the shuffled texts. Moreover,
the Esperanto text has more extreme values.  This suggests cascade model-like, with multiscale  time asymmetric features as finally written texts.  A discussion of the difference and complementarity of mapping into a LTS or FTS is presented. The FTS  $f(\alpha)$ curves are more opened than the LTS ones
\end{abstract}
 



 \maketitle



\section{Introduction} \label{sec-0}

The Hurst (or equivalently H\"older)  exponent  \cite{west}, measuring the so called self affinity of  signals, in short the roughness exponent,  can be generalized to some generalized fractal dimension $D$   \cite{west,BBM}.  However, multifractals \cite{halsey}    seem to better describe an object  through its evolving geometrical or structural features.  One has to recognize that there is some debate on whether   multifractality  exists  because of  finite size effects \cite{luxma}. The discussion on such a point should arise in some review article, outside the present paper. Let it be simply recalled that 
through a generator and from an initiator, one can easily produce a fractal object with a given dimension \cite{west}.  Note that to produce realistic and meaningful multifractal models is still a challenge \cite{schlumb}.    Next, one can ask "what to do with the knowledge that a dynamical object is a multifractal?"; even more: "How can this nonlinear  measure of  knowledge be useful?". Nevertheless, the first question is "Is there any multifractality evidence?".
 
Many authors have discussed the origin, characteristics, content, role of multifractals.  Let me point out to a pioneering experimental one \cite{schlumb}, a theoretical   \cite{DLA1}, a conceptual one   \cite{sornettePNAS}, and a few  so called applications \cite{mufractPNAS,PhD221mufrmusic,jafarimufractmusic}      in order  to set-up some wide perspective. Let us  also recall that one has  to obtain a $h(q)$ function which is a generalized Hurst or H\" older exponent or a $D(q)$ generalized dimension,  where $q$ represents the degree of some moment distribution of some time evolving variable. Subsequently one can obtain a $f(\alpha)$ spectrum, in which  $f(\alpha)$  is the distribution of
 the  exponent   $\alpha (\equiv \frac{d}{dq}[qh(q)])$    of the object.  

 A  written text can be considered 	as  a physical signal  \cite{eliade,martin}, because it can be decomposed through level thresholds which are like a set of characters taken from an alphabet.  As such, writings,  belong to the top level class of complexity \cite{saakian}.   One question immediately follows : are multifractals found in real texts? - a question already  raised in \cite{ebeling2} when studying  the  distribution  of letters in \textit{Moby Dick}; see also \cite{rodgers,antiqueira07,antiqueira09,fnote}.

  In \cite{ebeling,amit} it was claimed that long range order correlations (LROC) between words in texts express an author's ideas, and {\it in fine} even  consist in some author's signature \cite{rosso,holmes}.    
  Comparisons of $written$ texts $translated$ from one to another language  \cite{netwkIJMPC}, in particular  from the point of view of word LROC,  are of interest from the complexity point of view. The more so if the number of words in two languages is markedly different.    
In fact, since Shannon himself \cite{shannon}, writings and codings are of interest in statistical physics.  Writings are systems practically composed of a large number of internal components (the words, signs, and blanks in printed texts).

Texts, used here for investigating some {\it a priori} unknown structure, were chosen for their rather wide diffusion and incidentally being
 representative of a famous scientist, Lewis Carroll, i.e.\textit{ Alice in wonderland} (AWL) \cite{carollAWL,powers} and  \textit{Through a looking glass} (TLG)  \cite{carollTLG}. Knowing the mathematical quality of this author's mind,  one might expect  to find some special, unusual, unknown features of his texts. 
   Interestingly, a translation of  AWL  into Esperanto is available on  internet; here below, such a text will be referred to as ESP.  
   
    Having no  previous baseline for  such investigations,  the  three texts have been shuffled in order to  serve  as  base line.  This should allow  to check the robustness of the investigation methods and, if they exist, findings about multifractality of such written texts..

  In Sect. \ref{sec-1}, the data downloading and preliminary manipulations are explained. Next,
  the methodology is exposed: one can distinguish {\it frequency time series} (FTS)  from  {\it length time series} (LTS).  Different  techniques  exist  to investigate such supposedly multifractal signals. 
   Those are briefly recalled for  completeness.  Such techniques  are complementary; the presently used one sticks to the classical box counting method \cite{halsey}. The resulting data does not show any anomaly that would put into question the simplest method, and would request more fancy or advanced techniques.    
   
  More importantly, in the author's  opinion, one has to  remain within a statistical physics framework. In order to do so, one aim consists in searching for {\it correlations between fluctuations}, in the spirit of  the linear response theory \cite{LRTkubo,LRTkada}. Thus, the 12 time series  are transformed into  ''fluctuations'', i.e. series based on the  signs of the $"derivatives"$ of the texts (!), before calculating the multifractal features.
   
In Sect. \ref{sec-2},  
 the results  for the generalized Hurst exponent $h(q)$ and the  corresponding $f(\alpha)$ function \cite{halsey} are presented  and discussed.  In Sect. \ref{sec-3},  
 one comments about $indicators$,  i.e, the shape and extreme values of  $h(q)$, $\alpha$,  and  $f(\alpha)$ characterizing the texts. Those  suggest how to  analyze (dis)order and correlations, whence so called text  complexity, 
 along  cascade-like models \cite{Ottbook}, with multiscale  time asymmetric features.
 
  In Sect. \ref{sec-4},   a summary induces a conclusion.

 \section{Data and Methodology} \label{sec-1}
 
 The time series are made from a mapping of texts, here above mentioned,    downloaded from a freely available website   \cite{Gutenberg}. The chapter heads have first been removed before analysis. 
Three files are considered : (i)  the English version of  AWL, 
  - in short  AWL; (ii) its  translation into Esperanto, 
 -  in short ESP; and (iii)  and   the chronologically later  written (English) text TLG. 
 Note that even though the series  are to  be transformed, see below, the same notation is kept thereafter, referring as such to the original ($o$)  text without any ambiguity or to their shuffled version ($s$), i.e. AWLo, ..., TGLs.
  
    The shuffle algorithm is one found on Wikipedia. In brief, the first  data point is exchanged with some following  one, its location chosen from a generated random number. The second data point is exchanged with  some following one, chosen from another random number, etc.  The random number generator was checked to lead to a rather uniform distribution, for a number between 0 and 1. The algorithm was applied ten times on the texts to get the final shuffled texts hereby used for  analysis, comparison, and discussion.  In so doing,
the 6 documents have been transformed into 12 numerical
 one-dimensional
nonlinear maps in two ways \cite{LTSpanos} :
(i) by counting the number of occurrences  of each word in the whole  document, deducing  its frequency $f$.  The words are ranked accordingly, giving rank 1 to the most frequent word. Then, the text is "rewritten" into a series of numbers, such that  at each appearance of a word a number equal to its rank is  replacing the word.   Such a   series is called the frequency time series (FTS);
    (ii) by considering the length $l$ (number of letters) of a word. One records the word of length $l$ at each successive $"time"$   in the document, i.e. the first word is considered to be emitted at time $t $= 1, the second at time $t$ = 2, etc. A time series based on the amplitude $l(t)$ is so constructed.  It  is called a  length time series (LTS).
 
 Let it be mentioned that   punctuations and other typological signs are disregarded:  e.g.,  a "word" like $don't$ is considered as leading to $"don"$, - 3 letters, and $ "t"$, - 1 letter. The same goes on for  singular and plurals, giving two distinct words,  or verbs. For completeness, let it be mentioned that  the frequency, for example, of only lemmatized nouns or verbs could be studied \cite{0901.3291v1drzodz,netwkIJMPC}. Note that it should be obvious that the above mappings lead  to a continuous-like series, i.e. without blanks or gaps between words, now being numbers or a time index.

\begin{table}\label{tab-1}
\begin{center}
\begin{tabular}{|c||c|c|c|} \hline
&	AWLo  	&	ESPo 	&	TLGo 	 \\   \hline   \hline 
Number of   words	&	27342	&	25592	&	30601	\\  
Number of  different words	&	2958 	&	5368 	&	3205\\ 
Number of characters	&	144927	&	154445	&	164147	\\  
Number of ''sentences'' &	1633 	&	2016 	&	2059 \\  

Number of punctuation marks	&	4531 	&	4752 	&	4828 \\ 
 \hline
\end{tabular}
\end{center}
\caption{Basic statistical data for the three original texts of interest. The number of words gives the size of the "length  time series". The number of different words gives the size of the "frequency time series"}
 
\end{table}
 
  \begin{figure}  \begin{center}
 \hspace{-0.5cm}
 \includegraphics[width=3.0in] {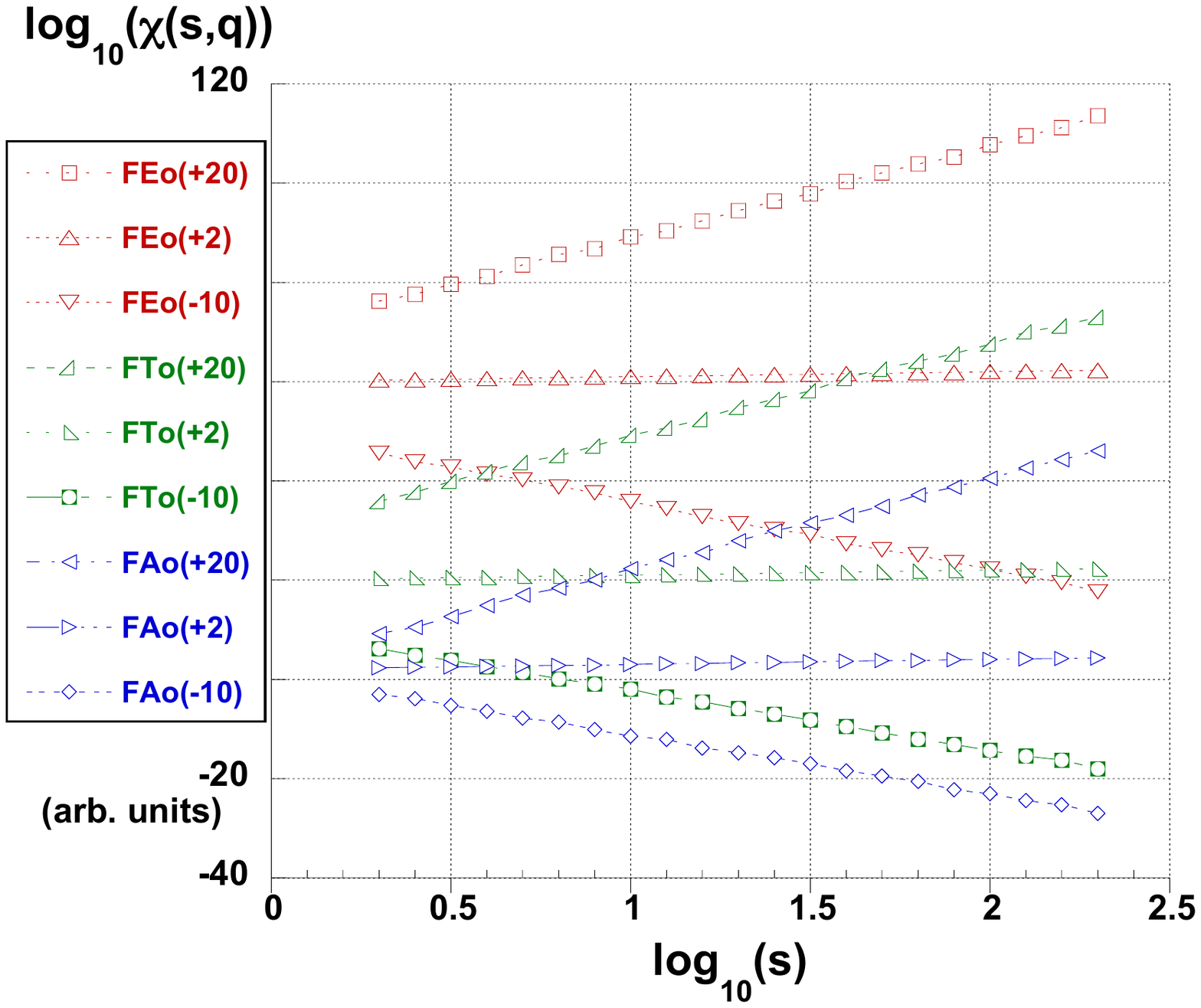}
 \vspace{-0.5cm}
 \includegraphics[width=3.0in] {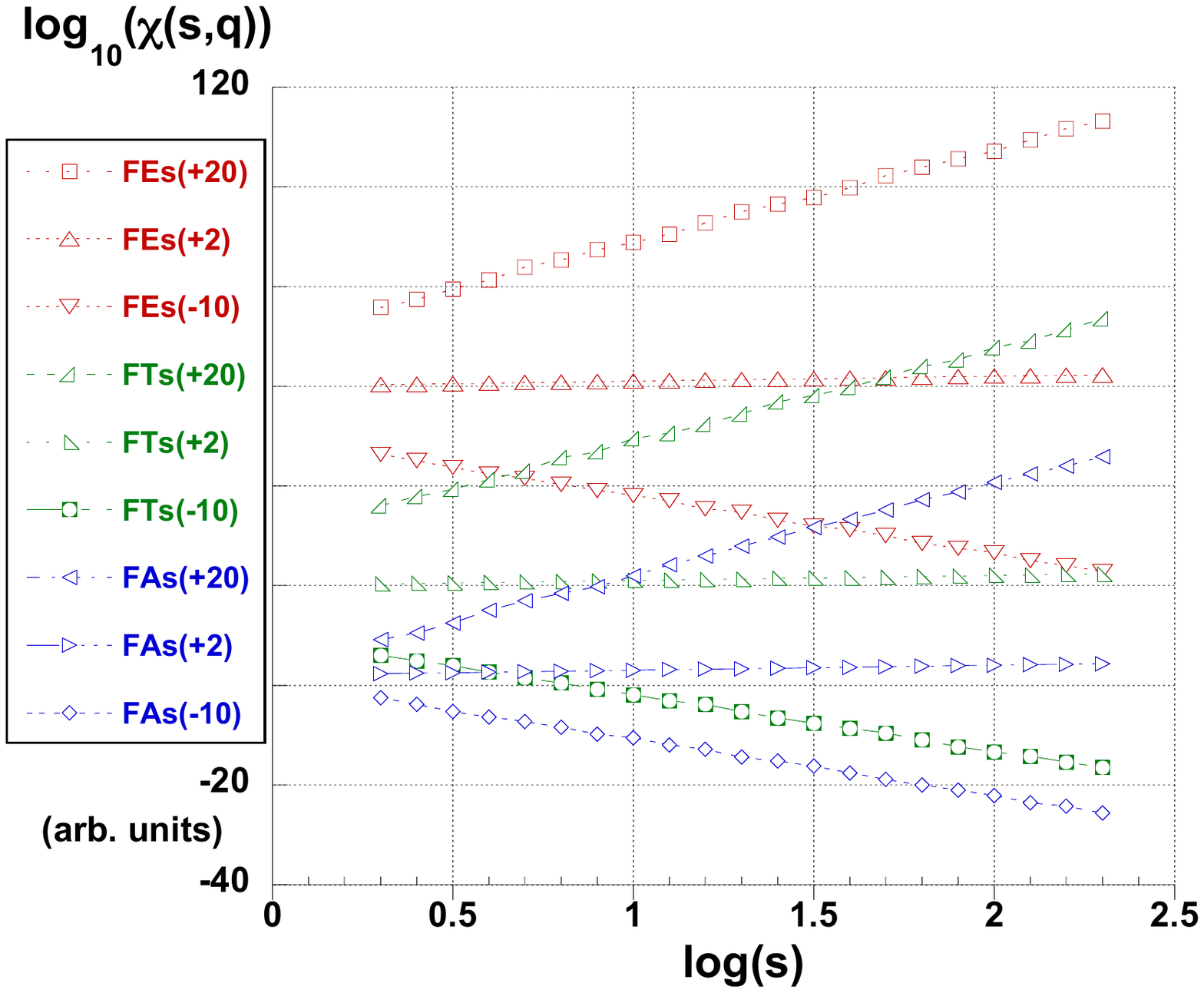}
 \vspace{-0.5cm}
 \caption{\label{fig-1ab}  The so called partition function $\chi(s,q)$ $vs$. $s$, the sub-series size in Eq. (1), on log-log plot graphs,  in order to obtain $\tau(q)$,   Eq. (3),   in the best possible power law regime, see text, and subsequently the generalized Hurst exponent $h(q)$, Eq. (4), or the generalized fractal dimension $D(q)$, Eq. (5), in the case of  FTS  for the (left)  original and (right)  shuffled  texts. Only 3 representative $q$-values (-10, +2, +20) in each case are shown for space savings. Obvious notations to understand the illustrating data are on  the left axis. In the display, the data has been arbitrarily displaced along the $y$-axis since only the slope from a linear fit is relevant} \end{center} \end{figure}

  \begin{figure}  \begin{center}
 \hspace{-0.5cm}
 \includegraphics[width=3.0in] {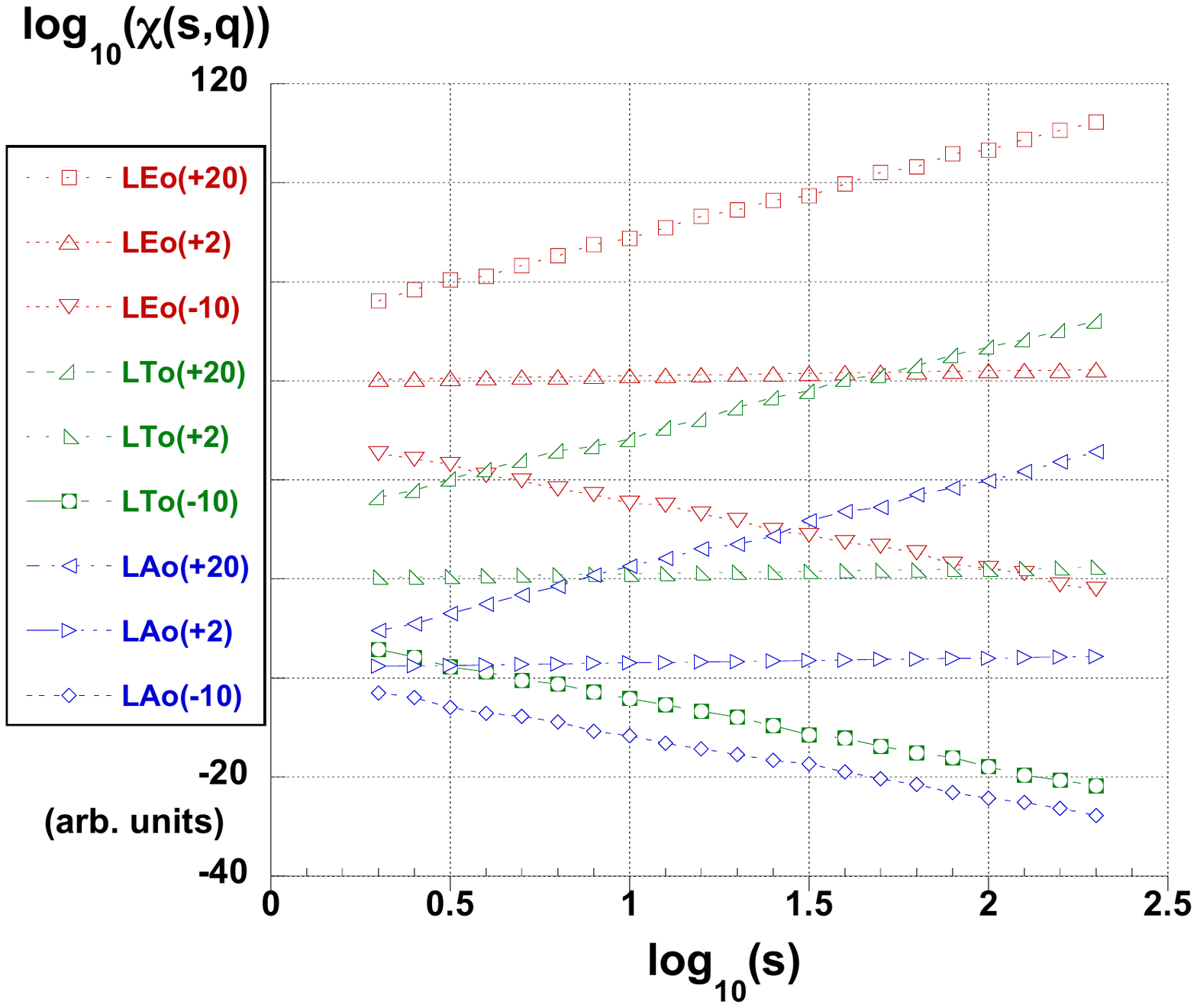}
 \vspace{-0.5cm}
 \includegraphics[width=3.0in] {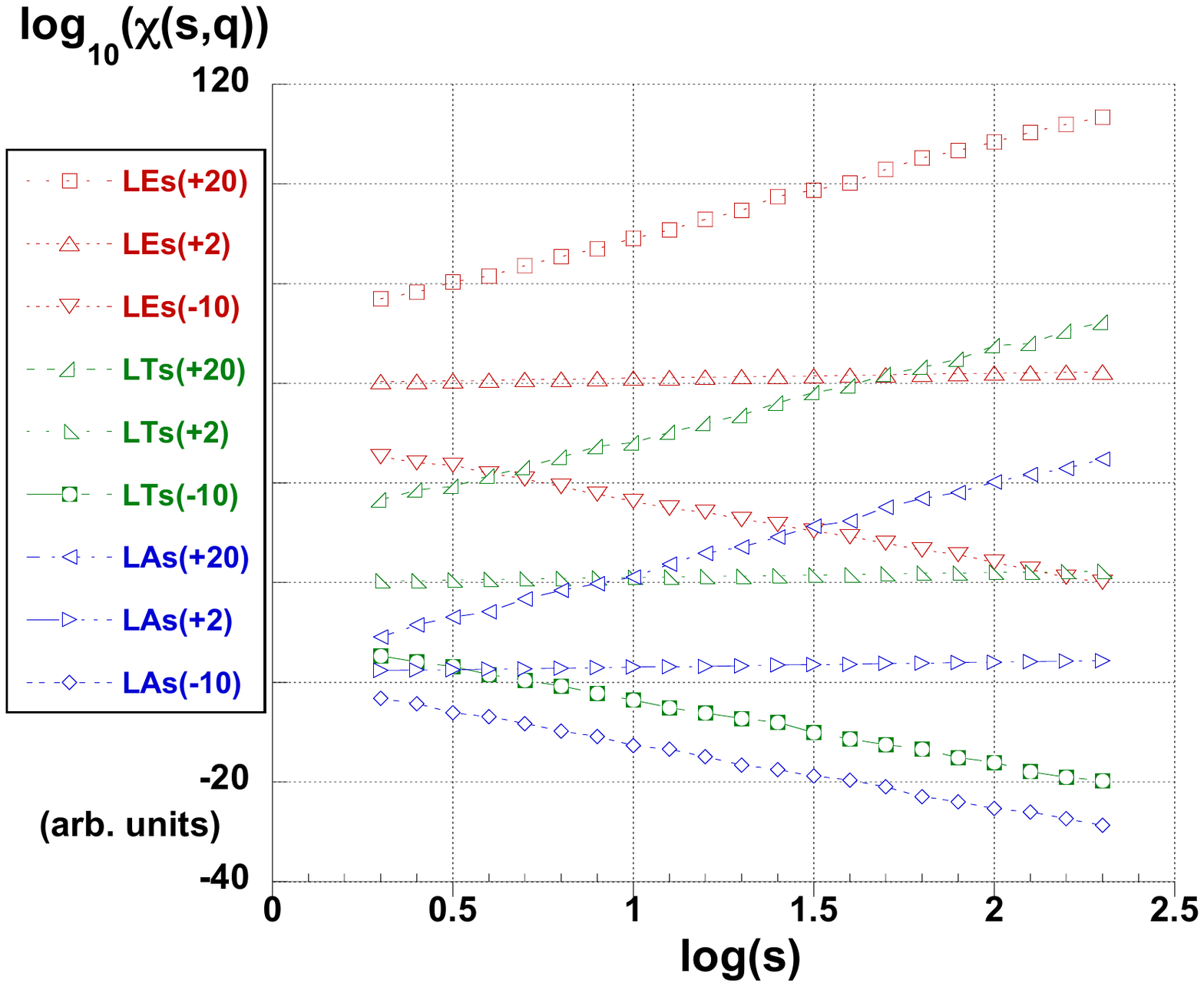}
 \vspace{-0.5cm}
 \caption{\label{fig-2ab} The so called partition function $\chi(s,q)$ $vs$. $s$, the sub-series size in Eq. (1), on log-log plot graphs,  in order to obtain $\tau(q)$,   Eq. (3),   in the best possible power law regime, see text, and subsequently the generalized Hurst exponent $h(q)$, Eq. (4), or the generalized fractal dimension $D(q)$, Eq. (5), in the case of  LTS  for the (left)  original and (right)  shuffled  texts. Only 3 representative $q$-values  (-10, +2, +20) in each case are shown for space savings. Obvious notations to understand the illustrating data are on the left axis. In the display, the data has been arbitrarily displaced along the $y$-axis since only the slope from a linear fit is relevant}
 \end{center} \end{figure}



\begin{figure}
\hspace{-0.5cm}
\includegraphics[width=3.0in]{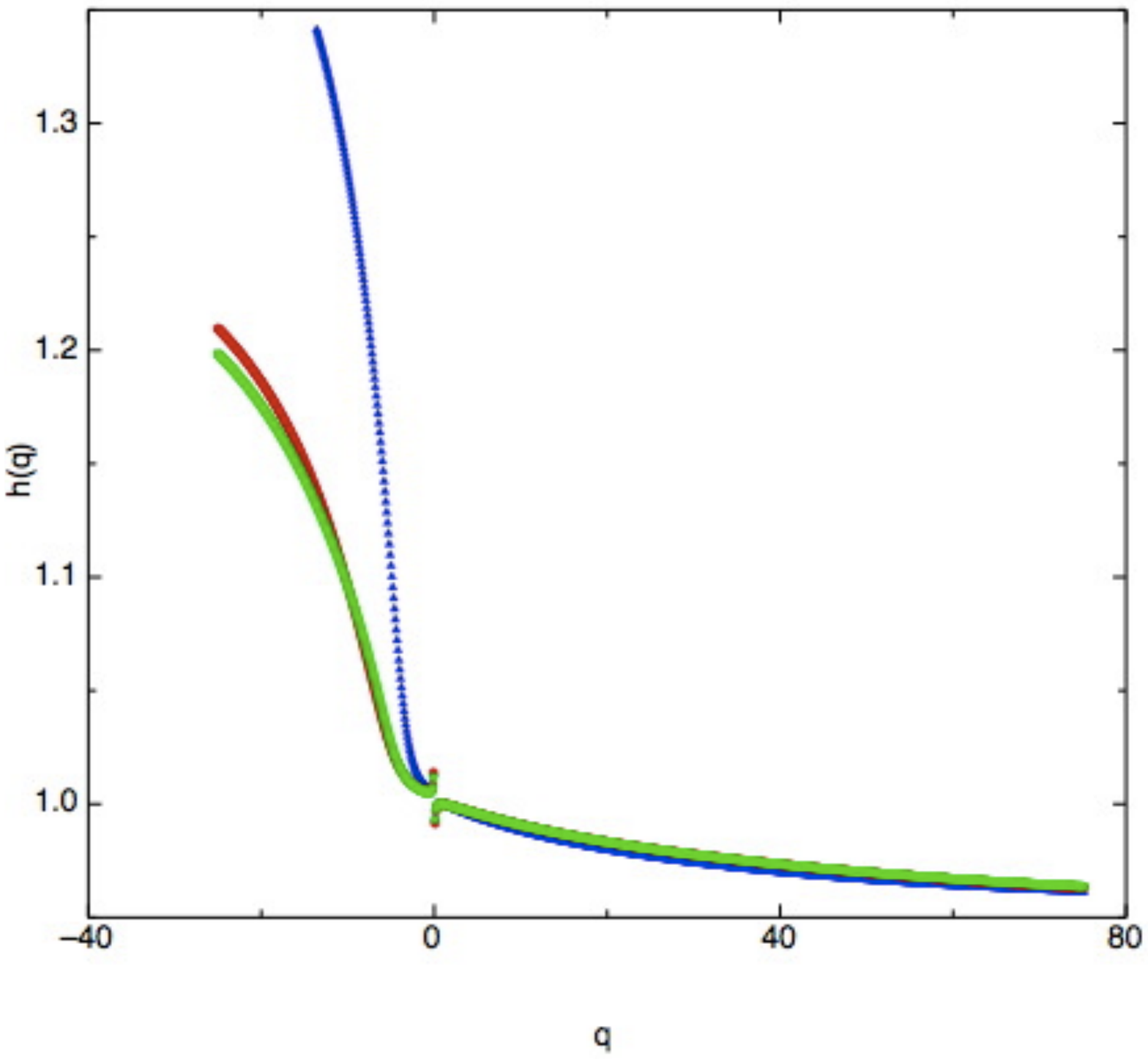}
\vspace{-0.4cm}
\includegraphics[width=3.0in]{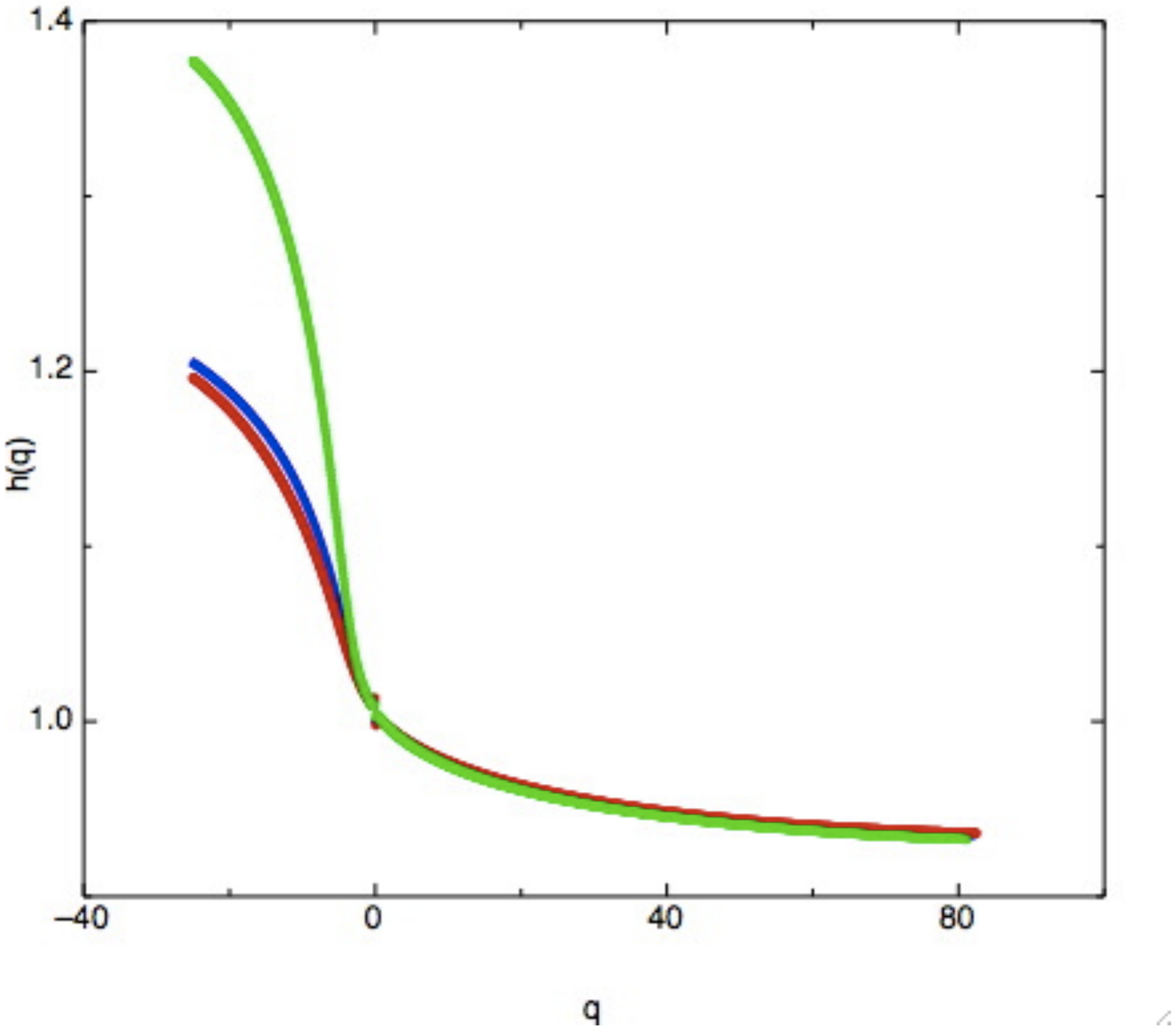}
\vspace{-0.15cm}
\caption{\label{fig-3}  Generalized Hurst exponent of three original texts analyzed through FTS (top) and  LTS (bottom) mapping: AWL: red, ESP: blue, TLG: green  dots}
\end{figure}

\begin{figure}
\hspace{-0.5cm}
\includegraphics[width=3.0in]{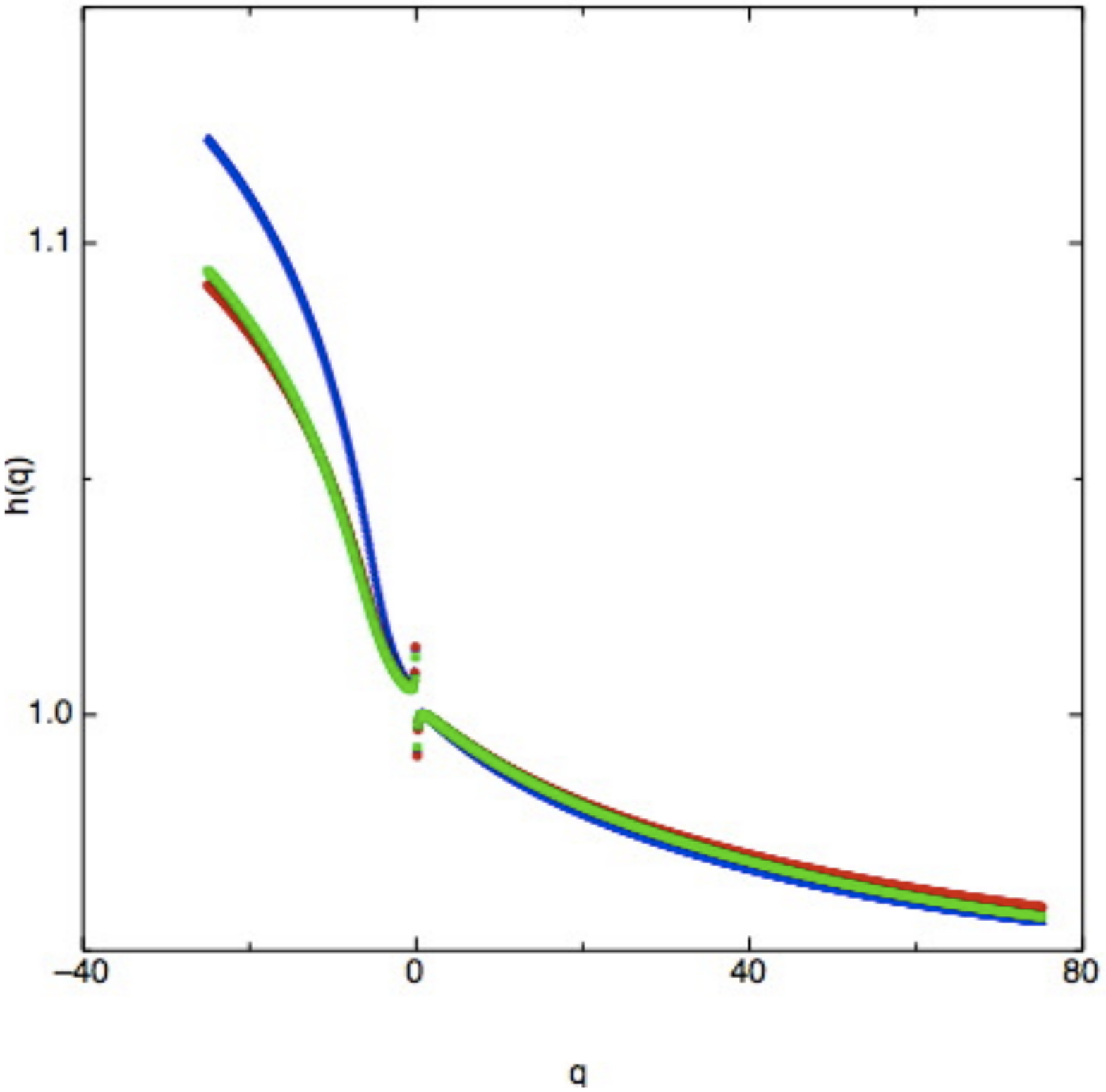}
\vspace{-0.4cm}
\includegraphics[width=3.0in]{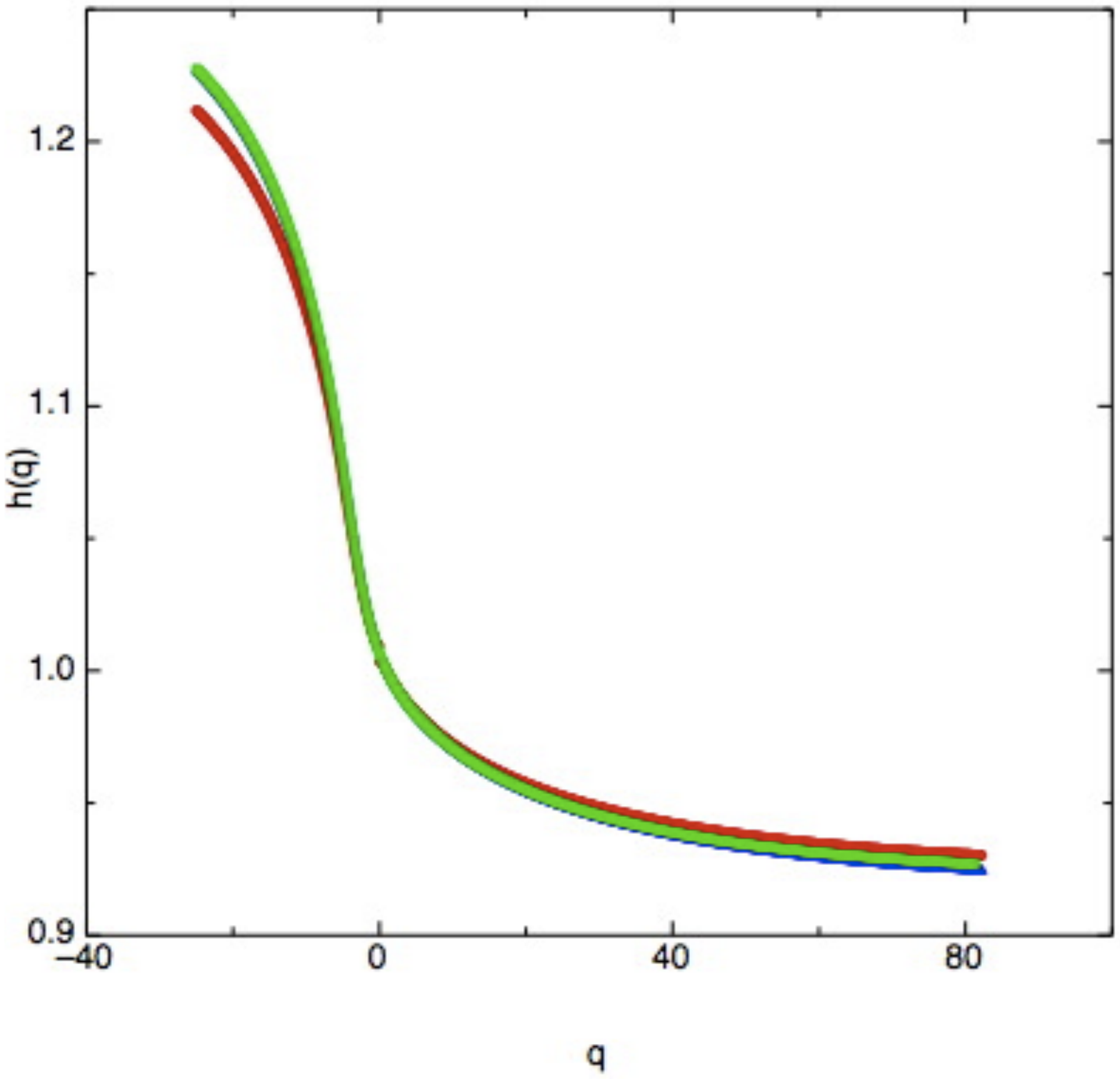}
\vspace{-0.15cm}
\caption{\label{fig-4} Generalized Hurst exponent of three shuffled texts analyzed through FTS (top) and  LTS (bottom) mapping: AWL: red, ESP: blue, TLG: green dots.}
\end{figure}

\begin{figure}  \begin{center}
\hspace{-1.5cm}
\includegraphics[width=3.6in] {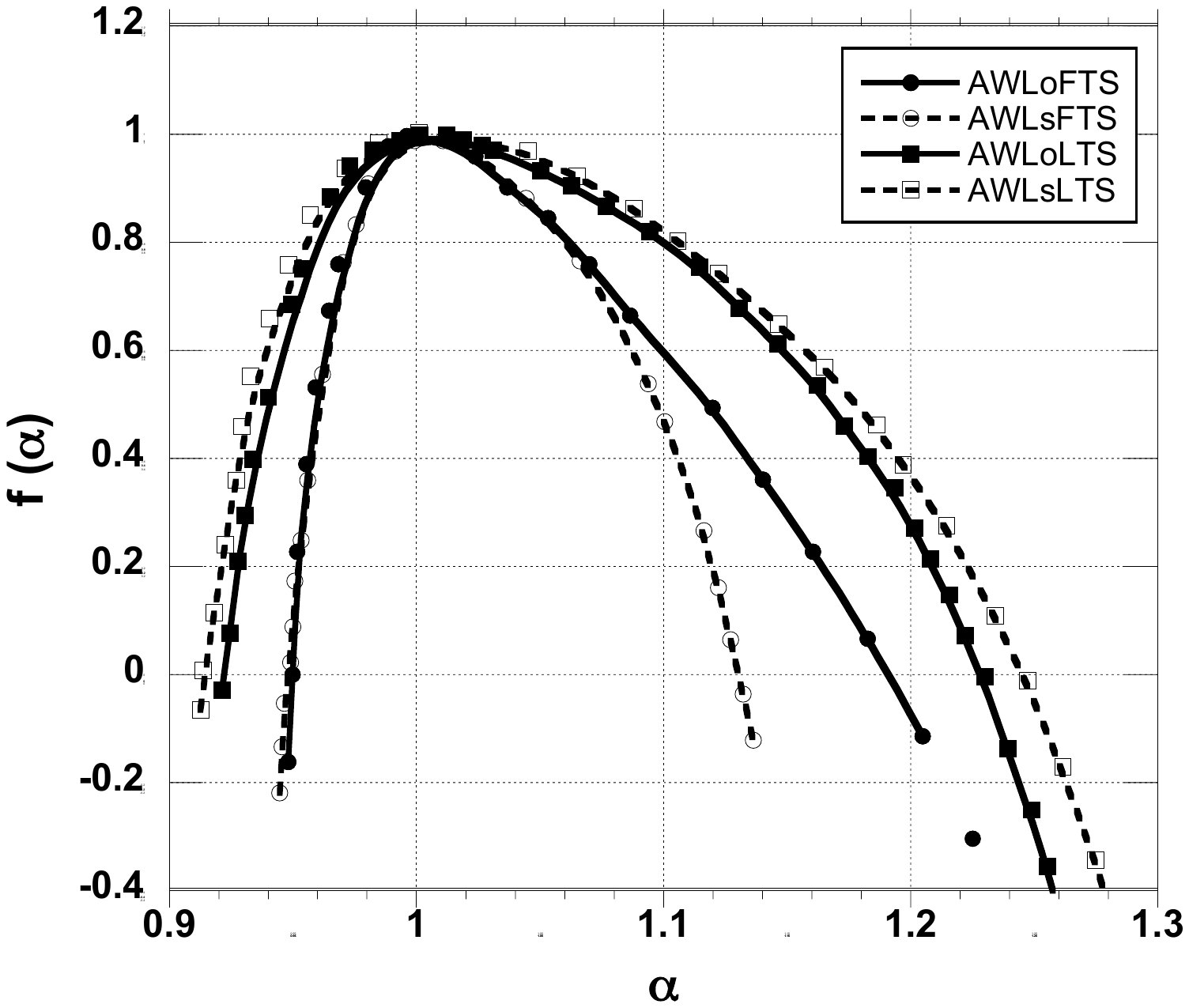}
\vspace{-3.5cm}
\caption{\label{fig-5faAWL}   $f(\alpha)$ for AWL, original (o)  or shuffled (s) text  along FTS  or  LTS  mapping}
\end{center} \end{figure}

\begin{figure}  \begin{center}
 \hspace{-1.5cm}
\includegraphics[width=3.6in] {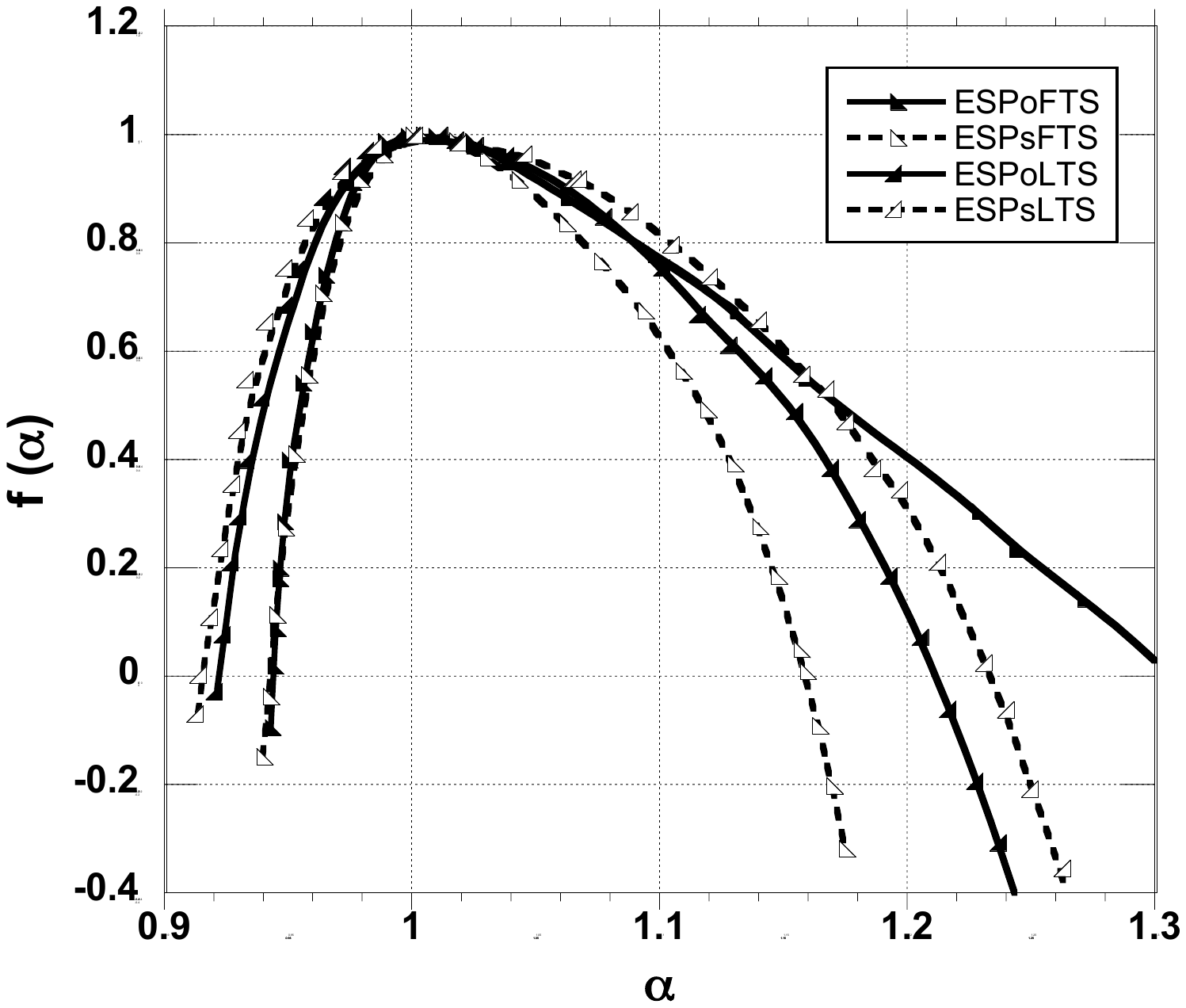}
\vspace{-3.5cm}
\caption{\label{fig-5faESP}   $f(\alpha)$ for ESP, original (o)  or shuffled (s) text  along FTS  or  LTS  mapping}
\end{center} \end{figure}

 \begin{figure}  \begin{center}
 \hspace{-1.5cm}
\includegraphics[width=3.6in]{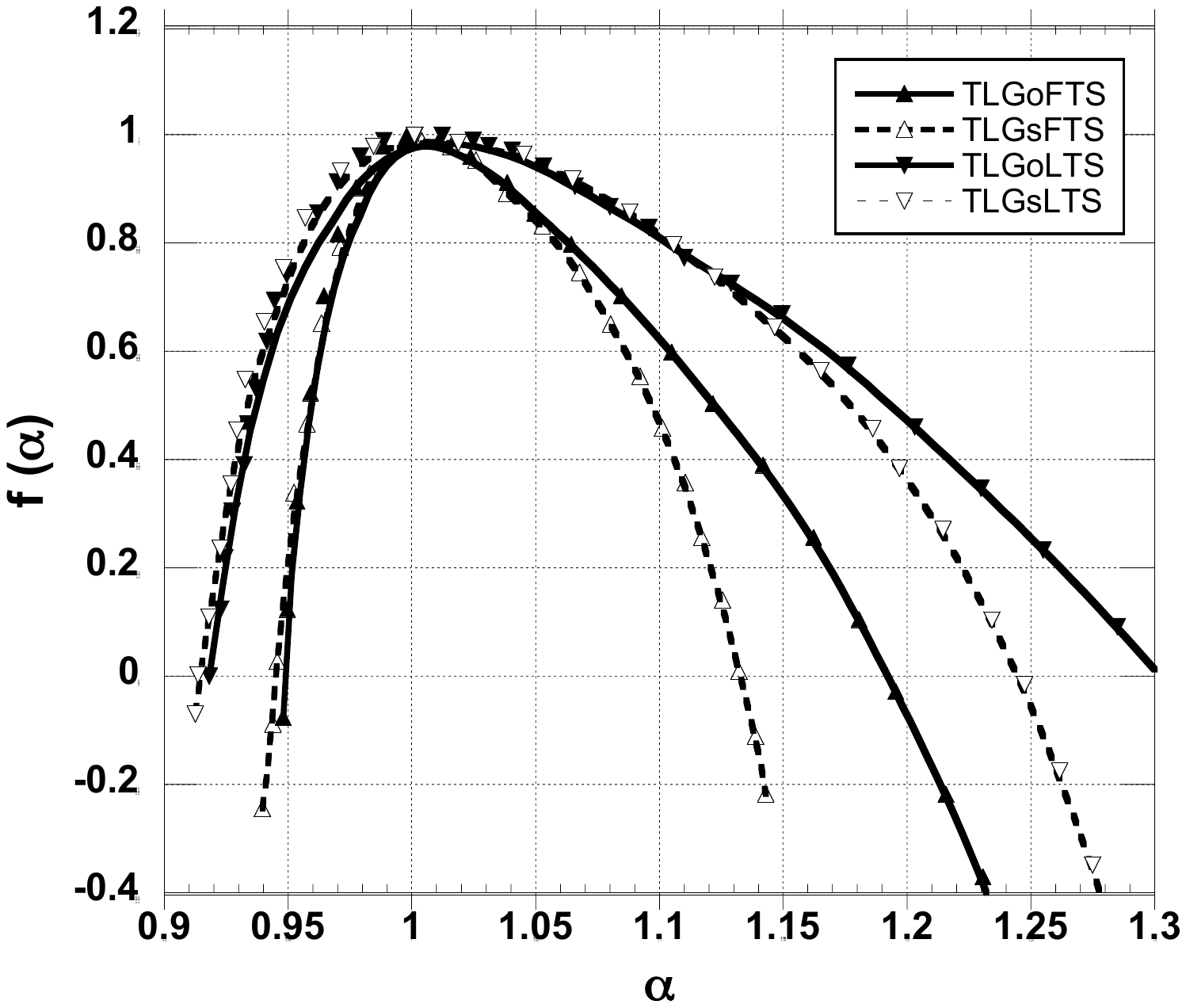}
 \vspace{-3.5cm}
\caption{\label{fig-5faTLG}   $f(\alpha)$ for TLG, original (o)  or shuffled (s) text  along FTS  or  LTS  mapping}
\end{center} \end{figure}

There are several techniques to demonstrate multifractality in time series, as nicely and recently reviewed in \cite{JWKencyclop} or by 
Schumann  and  
 Kantelhardt \cite{PhA390.11.2637JWK}. 
  Although the multiscaling features can be studied using different algorithms, each method  provides a complementary information about the complex structure of the time series.  
  
  One can be  analyzing either the statistics or the geometry, as well described 
  in \cite{PhA322.03.622turiel}. 

 A statistical approach consists of defining an appropriate intensive variable depending on a resolution parameter, then its statistical moments are calculated by averaging over an ensemble of realizations and at random base points. It is said that the variable is multifractal if those moments exhibit a power-law dependence in the resolution parameter. On the other hand, geometrical approaches \cite{NVMA297EPJB,NVMA300IJPC,KIMA327EPJB,KIMA394CBH,MAKI402CPC,KIMA391JAM}
try to assess a local power-law dependency on the resolution parameter for the same intensive variables at every particular point.
The geometrical approach is informative about the spatial localization of self-similar (fractal) structures, but leads to some   difficulty when having to justify the retrieval of scaling exponents.

The oldest multifractal analysis method is  the multifractal box counting (MF-BOX) technique \cite{halsey} which fails in presence of non-stationarities, such as trends. This deficiency led to the development of the wavelet transform modulus maxima (WTMM) method a generalized box counting approach based on a wavelet transform by Muzy-Bacry-Arneodo, as long ago as 1991  \cite{[23],muzyPRA45,[25],PhA309.02struzik,[35]}. Another approach to study multifractality in time series is the multifractal generalization of detrended fluctuation analysis (MF-DFA) of which Kantelhardt et al., on one hand, and Zunino et al., on the other hand  \cite{[36],PhA391.12.4100zunino,CSF41.09.233zunino,CSF42.09.2512queiros} are the most prolific represntatives. It  based on the traditional DFA \cite{[30]} or extensions \cite{NVMA299IJCAS,[31]}.

 Practically, MF-DFA is a less complicated and demands less presumption than the WTMM algorithm. For comparisons of these multifractal analysis methods, see \cite{[29],[36],[37],[38]}. Such comparisons indicate that MF-DFA is at least equivalent to WTMM, while an application of WTMM needs more care and yields spurious multifractality more often. In the present case, since there is no trend in such series, the simplest box counting technique is workable.  Thus,  the present study sticks to the classical box counting method \cite{halsey}. 

  \section{Results }  \label{sec-2}

 \subsection{Multifractal Analysis}

 The simplest type of multifractal analysis, based upon the standard partition function multifractal formalism \cite{halsey},  is summarized here below. 
 However, it is relevant to emphasize which variables are used in calculating the partition function. Since
 I want to stick to statistical physics ideas and methods, through, the usual "Linear Response Theory"  concepts \cite{LRTkubo,LRTkada} for calculating long range order features through  quantities, usually called susceptibilities,  it is useful to define the $fluctuations$ of interest before  calculating the correlations between  those. The most basic or primary fluctuations are in the derivative of a signal (or deviations from the mean, indeed).  In order to enhance the role of "fluctuations" in the time series, i.e. the text, each series is  transformed as follows, according to the most primary set of thresholds: if the length of a word in LTS (or its  frequency or rank in FTS) is smaller than the next one,  the former word gets a value = 2;  if it is greater, it gets   the value = 1; and 0 if both are equal. 
 The resulting series is called $ M_i$ ($1\leq i \leq N - 1 $). Next, each $ M_i $ is cut into $ N_s$ subseries of size $s$, where $N_s$ is the smallest integer in $ N/s$. The ordering starts from the beginning of the text, dropping out the last data points if necessary.  For either the original or shuffled text,  each FTS (or LTS) has  the same number of data points, $1 \leq i \leq N$.
  The number of words gives the size of the "length  time series". The number of different words gives the size of the "frequency time series". See such values and other informative data in Table 1.

Next, one calculates the probability
\begin{equation} \label{Peq}
P(s, \nu)  = \frac{\Sigma_{i=1}^{s}M_{(\nu-1)s+i}}{\Sigma_{\nu=1}^{N_s} \Sigma_{i=1}^{s}   M_{(\nu-1)s+i}}
\end{equation}
in "windows" of size $\nu$, for every $\nu$ and $s$.  Thereafter one calculates the so called partition function 
\begin{equation}\label{chieq}
\chi(s, q) = \Sigma_{\nu=1}^{N_s}  P(s, \nu)^q
\end{equation}
for each $s$ value.  A power law behavior  is expected
\begin{equation} \label{taueq}
\chi(s, q) \sim  s^{\tau(q)},
\end{equation}
where $\tau(q)$ plays the role of a partition function \cite{halsey}.
The generalized Hurst exponent, $h(q)$, is   obtained through
 \begin{equation}\label{hq}
 h(q) = \frac{1+\tau(q)}{q} .
 \end{equation}
 from   the best linear fit to Eq.({\ref{taueq}) on a log-log plot to get  $\tau(q)$.
 The generalized fractal dimension  $D(q)$ \cite{halsey} follows next:
\begin{equation} \label{Dqeq}
D(q) = \frac{\tau(q)}{q-1}.
\end{equation}
Let 
 \begin{equation}
 \alpha =  {d\tau(q)}/{dq}, 
 \end{equation}
from which, by inversion,  one obtains $q(\alpha)$ and $\tau(q(\alpha))$, whence the  $f(\alpha)$ function  \cite{12}  
 \begin{equation}
 f(\alpha) =  q \alpha - \tau(q),
 \end{equation}
as usual \cite{halsey}.

In the present work,   $\chi(s,q)$  has been calculated for  a very large $s$ range, i.e. between   2  and  5000, but the forthcoming below  reported data takes into account only the values  for  $2<s< 200$, i.e.  when $0.3<log_{10}$ $ s< 2.3$. In such a range,   the error bands are undistinguishable from the (mean of the) data; see   Figs. 3-7.  Moreover, the   $\tau(q)$ values must be measured in $s$ ranges where a power law, as in Eq.(3), is found.  Practically, one could do better in letting the extremal values of the $s$ interval be flexible, and, for example, let them be varied in each possible fit, but this is  much too time consuming for the final output, the more so if one attempts to cover a large set of $q$ values. The above mentioned extremal values were obtained, or rather "considered as acceptable", along the above criteria plus  some respect of computer time,  after many trial plots.
 
The   $\tau(q)$ values were calculated by a linear best fit on a log-log plot of  $\chi(s,q)$ $vs.$ $s$,  for $all$ (integer) $q$ values  such that -40 < $q$ < 80. Note that there are about 6 x 120 data set to fit.  Thus, the  number of $q$ values examined was reduced  to those such that   -35 $<q<$ 75, for FTS and to   -25 $<q<$ 80, for LTS. This allows one  to obtain smooth curves, see below, with  negligible error bars. 



Another  (technical)  comment, in advance of the reported results, in the following subsection,  is in order:  a too broad interval of $q$ might sometimes cast doubts on reported multifractality \cite{PhA387.08JiangZhou}.  
 It has been discussed that in the analysis of multifractality in turbulence or high-frequency financial data, the interesting moment orders $q$ should not be greater than 8 in order to make the partition function converge \cite{PhA387.08JiangZhou}.   However, as an example, the size of intraday high-frequency data is such that the moment order can be taken to be -120 $\leq$ $q$ $\leq$120 \cite{PhA387.08JiangZhou}.  In brief,
depending on the size of the time series,  the partition function  can be computed for rather large values of $q$,  if the convergence makes sense. In other words,  the error bars should become negligible or irrelevant  for the discussion purpose. As in other papers on multifractals  \cite{KIMAEPJB8,PhA308.02.518CBH,kitovasanantonio} or critical exponent search \cite{SSC65.88TbZn,MacDo} by the authors and co-workers, great care has thus been taken such that the here below presented data is  reliable both from  physics and   statistics criteria. No need to say that it takes much time  to do so and all steps are not recorded. The fit code  ($multifractalma.java$) is available from the author upon request.

 \subsection{$h(q)$ plots: Figs. \ref{fig-3}-\ref{fig-4}}

 For space savings, not all $\chi(s,q)$, Eq. (\ref{chieq}),  are shown here, as mentioned here above. However,  for a preliminary quantifying purpose, a summary of values, for $q$ = 2,  and its standard deviation,  found of the order of $10^{-3}$  are found in Table 2.  Recall that $q$ = 2 in fact corresponds to the standard DFA procedure. It is seen that the number of data points, i.e. the number of boxes of size $s$,  taken in order to estimate the slope of the straight line has some quite mild influence.  The latter might be a specific effect of time series based on written texts, or on the preliminary transformation of the time series into some sort of series of fluctuations. The matter has not been investigated further.

A few examples of plots of the partition function $\chi(s,q)$ $vs$. the sub-series size  $s$, see Eq. (1), are shown in Figs. \ref{fig-1ab}-\ref{fig-2ab}  on log-log graphs. As explained above, the $s$ range is chosen to be appropriate in order to obtain $\tau(q)$,   from Eq. (3).     In  each display, the raw data has been arbitrarily displaced along the $y$-axis, for good visualization purpose;  only the slope from a linear fit is relevant. It is already remarkable that the (positive or negative) slope values will be  of the same order of magnitude for the different but corresponding cases, either the original or shuffled series, with an expected evolution, as in many other studies. Also it is seen that there is hope for some possible distinction to be made between FTS and LTS cases depending on the original text.


From Eq. (4), the resulting $h(q)$ curves  of   the generalized  scaling,  Hurst, exponents, are given in Figs. \ref{fig-3}-\ref{fig-4}  for the various texts,  for $all$ (integer) $q$ values  such that -40 < $q$ < 80.  Observe that a marked numerical instability exists at $q=0$, - as usual, in fact, - better seen for the FTS than LTS .
For monofractal time series, $h(q)$ should be independent of $q$. 
 A multifractal structure is markedly observed, thus indicating that the scaling behaviors of small and large fluctuations are different.  It is known that the generalized Hurst exponent for negative   $q$ can be shown to describe the scaling of small fluctuations, - because the windows $\nu$, in Eq.(1),  with small variance   dominate for this $q$-range. In contrast, the windows $\nu$ with large variance have a stronger influence, - for positive $ q$.  Whence small fluctuations are usually characterized by larger scaling exponents than those related to large fluctuations, thereby inducing a Fermi or step function-like shape of $h(q)$.

\subsection{Note on $D(q)$  }

Obviously,
\begin{equation} \label{Dqhq}
D(q) = \frac{1}{q-1} (qh(q)-1).
\end{equation}

 First, observe the values of $h(2)$. For stationary signals,   $h(2)$ should coincide with the Hurst exponent, $H$, if the system is monofractal, and $D=2H-1$.  The  $h(2)$ values, as e.g. can be read from Figs.1-2,  are given in Table 3.  The values for the the shuffled texts lead to a doubtless fractal dimension = 1.    The slight deviations from unity for the original texts might be due to so called finite size effects.  Recall that the topology of the time series is a smooth line, without gaps. 
 
 Next, it can be deduced that  the generalized fractal dimension for the FTS  has a similar set of values for both english texts, decaying from $\sim$  1.2 to 1.0 for $q$ increasing but negative; $D(q)$ decays slowly for $q$ positive, barely reaching a value 0.95  for $q$ = 80. The value of $D(q)$ is much  greater along the negative $q$ axis, in particular for ESPo  but is identical to the other two for $q$  $\ge$ $0$. In LTS, the form of $D(q)$ is that to be expected and is similar to the FTS form.  
 
The shuffled texts  have remarkably similar $h(q)$, thus $D(q)$ values, both in range and variations, as those of the original texts, but the $D(q)$  values are closer to 1.0, - as could be expected. Very slightly  quantitative differences occur, - more markedly for the   EPSsFTS, see Fig. 2 (top), than for others. Along a Baeysian reasoning, these differences can be attributed to the finite size of the sample.

By the way,  
\cite{Vicsek}
 \begin{equation} 
 C_1 =    \left.{d\tau(q) \over dq} \right|_{q=1}
  \end{equation}  
  a measure of the intermittency lying in the signal $y(n)$, can be numerically estimated by measuring $\tau_q$ around $q=1$. In each case, the value of $C_1$ is close to unity (table of data not shown for space savings).
Some comment on the role/meaning of $C_1$, a sort of information entropy on  the structural complexity of a signal,  can be found  in Ref.\cite{MA4}.   

 \subsection{$f(\alpha)$  plots: Figs.  \ref{fig-5faAWL}-\ref{fig-5faTLG} }
 
 The $f(\alpha)$  spectra are shown in Figs. \ref{fig-5faAWL}- \ref{fig-5faTLG}. 
 Instead of presenting graphs based on FTS and LTS mappings, the data is presented for the three original texts and their shuffled counterpart. In so doing one can better compare for a given sample the methods and the subsequent results. 
 
 Before discussing the original texts/series, it can be observed that   the shuffling does not fully symmetrize the spectra. The rather finite size of these dynamical systems is likely the cause of such an imperfection. However, there is no doubt that all spectra  are markedly   non symmetric. This was at first found for DLA  simulations in  \cite{DLA1}, - with very high positive skewness, without much discussion. Note that for all series,  the FTS curves are wider than the LTS. In all cases also the original and its shuffled series lead to a $quasi$ identical  $f(\alpha)$  spectrum,   for  any $\alpha\le1$  and up to  $\alpha\simeq1.1$. 
 Above $\alpha\ge1.2$, some departure occurs, for several series, indicating a marked effect of large fluctuations.

\section{and some further Discussion }  \label{sec-3}
Let us stress linguistics-like  implications derived from the  above time series analysis of linguistics samples:
\begin{itemize} \item $h(q)$ and $D(q)$:
 In LTS, even though the form of $D(q)$ is that to be expected and is similar to the FTS form, it has to be stressed that the  AWLo and  ESPo are $very $ quantitatively similar, but {\it markedly differ} from  TLGo. This already indicates that one can observe a high  structural complexity  of the author 's style of writing through these  two books.  Moreover the multifractal analysis clearly shows that   a translation effect on the text style is much better observed through an FTS than an LTS.

Finally, it is fair to mention a reviewer remark: {\it the shuffled texts have remarkably similar $D(q)$ values. Does this mean that the multifractality is a 
distributional one and not due to non-linear correlations?} It could be the case indeed for the shuffled texts.

\item  $f(\alpha)$ : the  curve  rises very sharply: starting from negative values for $\alpha \le 1.0$, it reaches a maximum (=1.0) at 1.0, at the maximum so called box dimension, and decays less rapidly for $\alpha\ge 1$.
The not fully parabolic, to say the least, $f(\alpha)$ curve indicates non uniformity and strong LROC between long words and small words, - evidently arising from strong short range order correlations between these. In fact, the left  (right) hand side of the  $f(\alpha)$ curve  corresponds to fluctuations of the $q\ge0$ ($q\le0$)-correlation function. In other words, they correspond to correlated fluctuations in small (large) word distributions. 
 It  would be a nice conjecture that such distributions are personal  features of the vocabulary grasped by an author.  
 
 In so doing, the   the extremal $\alpha$ values, i.e. $ \alpha_- $ and  $ \alpha_+$ should be quantifying  the somewhat systemic way used by an author in his or her writings. These extreme values for  the 12 examined texts are given  in Table 3. 
 Observe that the Esperanto text differs from both English texts in such a consideration, - the English texts presenting the same . 

A short final note:  the Esperanto text curves behave differently from the English texts in FTS, though TLGo is different from the others in the LTS case. However, the shuffled texts $f(\alpha)$ spectra behave in a very similar way, both qualitatively $and$ quantitatively. I conjecture the effect to be due to the number of punctuation marks in such cases, see Table 1. Again, LROC and  the related structural complexity,    style  and creativity, are well exemplified. 

\end{itemize} 

\section{Conclusion}  \label{sec-4}
 
In summary, one has studied three samples, written texts, mapped as {\it in fine} 12 time series, due to introducing shuffled series as surrogate data for comparison. One can observe qualitative similarities between the original and shuffled texts and their translations, and quantitative differences.  
The English  texts  look more similar with each other than with respect to the Esperanto translation. The sharpness of $f(\alpha)$  indicates a high lack of uniformity of each text LROC.

The multifractal scheme   has been indicated to provide a measure of these correlations, thus a new indicator of a writer's style.  Of course, one might argue that  only text written by a single author, Lewis Carroll, are examined, not proving
  whether  the  so   obtained  $f(\alpha)$  
is text-dependent, writer-dependent, or both. 
That is why criteria suggested for estimating a text  {\it semantic complexity}
as if it is a time series are of interest.     It remains to be seen through more investigations  whether the $f(\alpha)$ curve and the  
cascade model hold true in other cases, and do  in general  characterize authors and/or texts, - and other time series.  Note that the multifractal  method should additionally be able to distinguish a natural
language signal from a computer code signal \cite{LTSpanos} and should help in improving translations by suggesting perfection criteria and indicators of  a translated text qualitative values, similar to those of the original one.   

Let it be re-emphasized the remarkable difference for the Esperanto text (Fig. 3a) with the English texts in the FTS analysis. Linguistics input should be searched at this level and is left for further discussion.  The origin of differences between TLG and AWL needs more work also at the linguistic level.

On the other hand, one physics conclusion arises from the above:  the existence of a multifractal spectrum found for the examined texts indicates a multiplicative process  in the usual statistical sense for the distribution of words length and frequency in the text considered as a time series.  Thus linguistic signals may be considered indeed as the manifestation of a complex system of high dimensionality, different from random signals
or from systems of low dimensionality such as the  financial and geophysical (climate) signals.   In so doing one can consider the behavior of the atypical $f(\alpha)$ curve as originating from a binomial multiplicative cascade process  as in fully developed turbulence \cite{Ottbook},   here for short and long words, on a support [0,1].  

Extensions to higher dimensions, e.g.  in image recognition \cite{0801.2501v1MonaLisa}   or in hypertext studies are thus quite possible.    In relation  to these remarks, work on fractal analysis of   paintings should be mentioned \cite{taylorpollock,0801.2501v1MonaLisa}, on handwriting \cite{handwriting}  and on japanese garden patterns \cite{Japanesegardens} to indicate directions for further research.

\hspace{2.8cm}

{\bf Acknowledgements} \\

Comments by reviewers have surely improved this paper. I would like also   to thank N.K. Vitanov,  G. Rotundo, and  A. Scharnhorst  for, as usual, fruitful comments and discussions, and   J.  Gillet  for much help with the data acquisition and its analysis.  
  Thanks also to the COST Action MP0801  which provided some financial support  to stay at BAS, Sofia, BG,   U. Tuscia, Viterbo,  IT, and  DANS-KNAW, Amsterdam, NL,  through the STSM 5378, STSM 6698, and STSM 9874 respectively .

 \hspace{1.8cm}


\hspace{1.8cm}

\newpage

 \hspace{1.8cm}

\begin{table}[ht]
\label{tab-1DFA}
 \caption{ Characteristic  slope values,  for $q=2$, for the original (o) and  shuffled (s) texts, according to the type of series (FTS or LTS) so examined 
 }
 \begin{footnotesize}
 \begin{center}  
\begin{tabular}{@{\vrule height 10.5pt depth4pt  width0pt}|c|  c c| c c| c c| } \hline 
 \centering{Original  texts} & $0-200$& $std.dev.$ & $200-5000$ &$std.dev..$& $0-5000$&$std.dev.$ \\ \hline 
AWLoFTS  &0.491&2E-3&0.561&2E-3&0.561&2E-3  \\ 
	ESPoFTS &0.519&2E-3&0.544&1E-3&0.545&1E-3	\\ 
	TLGoFTS   &0.501&2E-3&0.777&3E-3&0.774&3E-3  \\ \hline
AWLoLTS &0.538&2E-3&0.686&1E-3&0.684&1E-3 	 \\ 
	 ESPoLTS  &0.516&2E-3&0.619&2E-3&0.620&2E-3	\\ 
	 TLGoLTS&0.531&2E-3&0.560&1E-3&0.560&2E-3    \\ \hline\hline
 \centering{Shuffled  texts} & $0-200$&$std.dev.$ & $200-5000$ &$std.dev.$& $0-5000$&$std.dev.$
 \\ \hline 
AWLsFTS&0.525&1E-3&0.534&1E-3&0.533&1E-3  \\ 
	ESPsFTS &0.518&1E-3&0.474&1E-3&0.478&1E-3   \\ 
	TLGsFTS   &0.524&1E-3&0.480&1E-3&0.480&1E-3	 \\ \hline 
 AWLsLTS&0.461&2E-3&0.584&1E-3&0.581&1E-3  \\ 
	ESPsLTS  &0.519&4E-3&0.507&1E-3&0.506&1E-3   \\ 
	TLGsLTS  &0.504&3E-3&0.587&1E-3&0.584&1E-3	 \\ \hline 
\end{tabular}
 \end{center}
 \end{footnotesize}
\end{table}

\begin{table}[h]
\label{tab-3}
 \caption{ Characteristic $h(q=2)$,  $\alpha_-$ and $\alpha_+$ values, see Figs. 1-5,  for the original, translated and  shuffled texts, according to the type of series (FTS or LTS) so examined
 }
 \begin{footnotesize}
 \begin{center}  
\begin{tabular}{@{\vrule height 10.5pt depth4pt  width0pt}  |c| |c| c  c| } \hline 
 \centering{Original Texts } &$h(2)$& $\alpha_-$& $\alpha_+$   \\ \hline 
 AWLoFTS &$0.997$   &0.95&1.19  \\ 
 ESPoFTS  &$0.997$   &0.94&1.30 	\\ 
 TLGoFTs &$0.997$    &0.95&1.19   \\ \hline
 AWLoLTS&$0.994$    &0.92&1.23 	 \\ 
 ESPoLTS&$0.994$    &0.92&1.21  	\\ 
 TLGoLTS&$0.994$    &0.92&1.34   \\ \hline\hline
 \centering{Shuffled  Texts}&$h(2)$   & $\alpha_-$& $\alpha_+$ 	 \\ \hline 
 AWLsFTS&1.0 &0.95&1.13  \\ 
 AESPsFTS &1.0 &0.96&1.16    \\ 
 TLGsFTS& 1.0   &0.94&1.13 	 \\ \hline 
 AWLsLTS&0.999  &0.91&1.25   \\ 
 ESPsLTS&0.999  &0.92&1.24   \\ 
TLGsLTS & 0.999   &0.91&1.25 	 \\ \hline 
\end{tabular}
 \end{center}
 \end{footnotesize}
\end{table}


\begin{thebibliography}{0} 


 \bibitem{west} B.J. West and W. Deering,  {\it  The lure of modern science: fractal thinking} (World Sci., River Edge, NJ, 1995);  ibid.,  Phys. Rep. {\bf 246}, 1 (1994).
 
\bibitem{BBM}  B.B. Mandelbrot,   {\it The Fractal Geometry of Nature}  (W.H. Freeman, New York, 1982).

 \bibitem{halsey} T.C. Halsey, M.H. Jensen, L.P. Kadanoff, I. Procaccia, and B.I. Shraiman, 
Phys. Rev. A {\bf 33}, 1141 (1986).

\bibitem{luxma} Th. Lux and M. Ausloos, ÒMarket Fluctuations I : Scaling, Multi-scaling and their Possible OriginsÓ, in  {\it ÒThe Science of Disaster: Scaling Laws Governing Weather, Body, Stock-Market DynamicsÓ}, A. Bunde, J. Kropp, and  H.-J. Schellnh\" uber, Eds. (Springer Verlag, Berlin, 2002) pp. 377-413 

   \bibitem{schlumb} J. Nittmann, H. E. Stanley, E. Touboul, and G. Daccord,
   Phys. Rev. Lett.  {\bf 58}, 619 (1987).
       
\bibitem{DLA1} S. Schwarzer, J. Lee, A. Bunde, S. Havlin, H. E. Roman, and H. E. Stanley, 
 Phys. Rev. Lett.  {\bf 65}, 603  (1990).
  
  \bibitem{sornettePNAS} 
D. Sornette,  A. B. Davis, K. Ide, K. R. Vixie, V. Pisarenko, and J. R. Kamm,
Proc. Natl. Acad. Sci. USA. 
{\bf 104},   6562 (2007).

  \bibitem{mufractPNAS} A.L.. Goldberger,  L.A. Amaral, J.M. Hausdorff, P.Ch. Ivanov,  C.K. Peng, and H.E. Stanley,
  Proc. Natl. Acad. Sci. USA. 
  {\bf 99 Suppl. 1},    2466 (2002).
  
    \bibitem{PhD221mufrmusic}  Zhi-Yuan Su and Tzuyin Wu, Physica D {\bf 221}, 188 (2006).

  \bibitem{jafarimufractmusic}  G. R. Jafari, P. Pedram, and L. Hedayatifar,  J. Stat. Mech. (2007) P04012.
   
   
  \bibitem{eliade}  M. Eliade, W.
  R. Trask,  and J.
  Z. Smith,  {\it  The Myth of the Eternal Return: Cosmos and History} (Princeton Classic Editions, Princeton UP, Princeton, 1971).
  
  \bibitem{martin}   J. Martin,    {\it English text: System and structure} (John Benjamins, Amsterdam, 1992).

\bibitem{saakian} D. B. Saakian,
Phys. Rev. E  {\bf 71}, 016126 (2005).

 \bibitem{ebeling2} A.N. Pavlov,  W. Ebeling, L. Molgedey, A. R. Ziganshin, and V. S. Anishenko, 
   Physica A  {\bf 300},  310  (2001).                

\bibitem{rodgers} In fact, it has been  recently proposed that writings can be thought as being mapped on networks: A. P. Masucci and G. J. Rodgers,
Phys. Rev. E  {\bf 74}, 026102 (2006).
   
   \bibitem{antiqueira07}   L. Antiqueira, M. G. V. Nunes, O. N. Oliveira Jr., and L. F. Costa,  
 Physica A {\bf 373}, 811 
(2007).

   \bibitem{antiqueira09}  L. Antiqueira, O. N. Oliveira Jr.,  L. F. Costa, and M. G. V. Nunes,  
Inform.  Sciences {\bf  179},   584 (2009).  

\bibitem{fnote} These have fractal properties; 
most usually they should be multifractals; one can thus imagine/consider/argue  that a text   is a form of partially self-organized   network  of words due to grammatical and style constraints, thus should present multifractal features.

  
  
   
    \bibitem{ebeling} W.  Ebeling and A.  Neiman,
   Physica A {\bf  215}, 233 (1995).   
   
       \bibitem{amit} M. Amit, Y. Shmerler, E. Eisenberg, M. Abraham, and N. Shnerb, 
  Fractals {\bf 2}, 7  (1994).  
 
\bibitem{rosso} O.  A. Rosso and H. Craig Pablo Moscato,
 Physica A {\bf 388}, 916 (2009).

\bibitem{holmes} D.  I. Holmes,
Comput. Humanities {\bf 28},  87 (1994).

   \bibitem{netwkIJMPC} D.R. Amancio, L. Antiqueira. T.A.S. Pardo, L. da F. Costa, O.N. Oliveira Jr.,  and M. G. V. Nunes,  
      Int. J. Mod. Phys. C {\bf  19}, 583  (2008).
      
       \bibitem{shannon} C. Shannon,  
       	Bell. Syst. Tech. J. {\bf 27},  379 (1948); $ ibid.$, Bell. Syst. Tech. J. {\bf 27},  623  (1948); see also 
		  $ ibid.$,  Bell Syst. Tech. J. {\bf 30},  50 (1951).
      
 \bibitem{carollAWL} L.W. Carroll,  {\it Alice's Adventures  in Wonderland } (Macmillan, New York, 1865);  

see $http://www.gutenberg.org/etext/11$.

\bibitem{powers}  Previous work on  the english AWL version can be mentioned but pertains to a mere Zipf analysis.; D.M.W. Powers, Applications and explanations of Zipf's laws, in {\it New Methods in Language Processing and Computational natural Language Learning } D.M.W. Powers (ed.) 
(ACL, 1998) pp 151-160.
  
  \bibitem{carollTLG} L.W. Carroll,  {\it Through the Looking Glass and What Alice Found There} (Macmillan, New York, 1871);  

see $http://www.gutenberg.org/etext/12$.

   \bibitem{LRTkubo}  R. Kubo,   
    Rep. Progr. Phys. {\bf 29}, 255 (1966).

\bibitem{LRTkada}  L. P. Kadanoff and P. C. Martin, 
Ann. Phys. -New York  {\bf 24}, 419   (1963);  reprinted Ann. Phys.-New York   {\bf 281}, 800 (2000).
            
\bibitem{Ottbook} 
E. Ott,  {\it Chaos in Dynamical Systems}, 2nd Ed. (Cambridge U.P., Cambridge, 2002) ch. 9.

 \bibitem{Gutenberg}  Project Gutenberg (National Clearinghouse for Machine Readable Texts)
$http://www.gutenberg.org$.
  
  \bibitem{LTSpanos} K. Kosmidis, A. Kalampokis,  and P.  Argyrakis,
 Physica A {\bf   370},  808  (2006).
 
\bibitem{0901.3291v1drzodz}  S. Dro\.zd\. z, J. Kwapie\' n, and A. Orczyk, Approaching the linguistic complexity,
 arXiv:0901.3291 (2009);  A. Orczyk, Master thesis, University of Science and Technology AGH, Krak\'ow, Poland (2008);  J J. Kwapie\' n,  S.	Dro\.zd\.z,  and A. Orczyk, Acta Phys. Pol. A  {\bf  117},  716 2010).
  
 \bibitem{JWKencyclop}  J.W. Kantelhardt, Fractal and multifractal time series, in: R.A. Meyers (Ed.), Encyclopedia of Complexity and Systems Science, (Springer-Verlag, Berlin, 2009).
 
    \bibitem{PhA390.11.2637JWK} A. Y. Schumann  and   J. W.  Kantelhardt,
Physica A {\bf 390},   2637 (2011).

    \bibitem{PhA322.03.622turiel} A. Turiel    and C.J.
P\' erez-Vicente, Physica A {\bf 322},  629 (2003).

   \bibitem{NVMA297EPJB} N. Vandewalle and M. Ausloos, 
   	Eur. J. Phys. B  {\bf 4}, 257 (1998).
   
      \bibitem{NVMA300IJPC} N. Vandewalle and M. Ausloos, 
      	 Int. J. Phys. C {\bf  9}, 711 (1998).
      
         \bibitem{KIMA327EPJB} K. Ivanova and M. Ausloos,
         	Eur. Phys. J. B  {\bf  8},  665  (1999); Err.  {\bf  12}, 613 (1999).
         
            \bibitem{KIMA394CBH} K. Ivanova, H.N. Shirer, E.E. Clothiaux, N. Kitova, M.A. Mikhalev, T.P. Ackerman, and M. Ausloos, 
            	Physica A  {\bf  308}, 518 (2002).  
            
               \bibitem{MAKI402CPC} M. Ausloos and K. Ivanova,  
               Comp. Phys. Commun. {\bf 147}, 582 (2002).
               
                  \bibitem{KIMA391JAM} K. Ivanova, E.E. Clothiaux, H.N. Shirer, T.P. Ackerman, J.C. Liljegren, and M. Ausloos, 
                  J. Appl. Meteorology  {\bf  41}, 56  (2002).

 \bibitem{[23]} J.F.Muzy, E.Bacry, and A. Arneodo, Phys. Rev. Lett.  {\bf  67}, 3515 (1991). 
 
    \bibitem{muzyPRA45} J.F. Muzy,  B. Pouligny, E. Freysz, F. Argoul, and A. Arneodo.
	Phys. Rev. A {\bf  45} 8961 
(1992).

   \bibitem{[35]} J.F. Muzy, E. Bacry, and A. Arneodo,  
   Internat. J. Bifur. Chaos  {\bf 4},  245 (1994). 

  \bibitem{[25]} A. Arneodo, E. Bacry, P.V. Graves, and J.F. Muzy,  
  Phys. Rev. Lett. {\bf  74}, 
 3293 (1995).
  
   \bibitem{PhA309.02struzik} Z. R. Struzik, and A. P.J.M. Siebes, Physica A {\bf  309}, 388 (2002).

  

 \bibitem{[36]} J.W. Kantelhardt, S.A. Zschiegner, E. Koscielny-Bunde, S. Havlin, A. Bunde, and  H.E. Stanley, 
   	Physica A {\bf 316},   87  (2002).
 
 
  \bibitem{PhA391.12.4100zunino} 
   D. Gulich and L. Zunino, Physica A {\bf  391}, 4100 (2012).
  
   \bibitem{CSF41.09.233zunino} L. Zunino, A. Figliola, B.M. Tabak, D.G. P\' erez, M. Garavaglia, and O.A. Rosso, 
   	Chaos, Solitons \& Fractals {\bf 41}, 2331  (2009).
   
 \bibitem{CSF42.09.2512queiros}  J. de Souza  and S. M. Duarte Queir\' os,  Chaos, Solitons \& Fractals
{\bf 42},   2512 (2009). 

 \bibitem{[30]} C.-K. Peng, S.V. Buldyrev, S. Havlin, M. Simons, H.E. Stanley, and A.L. Goldberger,  Phys. Rev. E  {\bf 49}, 1685  (1994).
 
 \bibitem{NVMA299IJCAS} N. Vandewalle and M. Ausloos, 
 	Int. J. Comput. Anticipat. Syst. {\bf 1}, 342 (1998).
 
 
 \bibitem{[31]}  X.-Y. Qian, G.-F. Gu, and W.-X. Zhou, 
 	Physica A {\bf 390}, 4388 (2011).

  \bibitem{[29]} J.W. Kantelhardt, D. Rybski, S.A. Zschiegner, P. Braun, E. Koscielny-Bunde, V. Livina, S. Havlin, and  A. Bunde, 
	Physica A {\bf 330},  240 (2003).

   
    \bibitem{[37]}
    P. O\' swiecimka, J. Kwapie\' n,  and S. Dro\.zd\. z, 
    	Phys. Rev. E {\bf 74}, 016103 (2006). 
    
     \bibitem{[38]} A. Turiel, C.J. Perez-Vicente, and J. Grazzini, 
      	J. Comput. Phys. {\bf 216},   36 (2006).
	
  \bibitem{12} A. Chhabra and R.V.  Jensen, 
Phys. Rev. Lett.  {\bf 62},  1327 (1989).
   
  \bibitem{PhA387.08JiangZhou}
Z.-Q. Jiang and W.-X. Zhou, Physica A {\bf  387}, 3605 (2008).

\bibitem{KIMAEPJB8} K. Ivanova  and M. Ausloos,  
Eur. Phys. J. B {\bf 8}, 665 (1999). 

\bibitem{PhA308.02.518CBH} K. Ivanova, H.N. Shirer, E.E. Clothiaux, N. Kitova, M.A. Mikhalev, T.P. Ackerman, and M. Ausloos, 
 Physica A   {\bf 308},  518 (2002).



\bibitem{kitovasanantonio} N. Kitova,  M. A. Mikhalev, K. Ivanova, M. Ausloos, and T. P. Ackerman,
 Fourteenth ARM Science Team Meeting Proceedings, Albuquerque, New Mexico, March 22-26, 2004,
$http://www.arm.gov/publications/proceedings/$ $conf14/$
 $extended_-abs/kitova-n.pdf$.
 
\bibitem{SSC65.88TbZn} M.M.  Amado, R.P.  Pinto, J.M.  Moreira, M.E.  Braga, J.B.  Sousa, P.  Morin, P.  Clippe and M.  Ausloos,  
 Solid State Commun. {\bf 65}, 1429    (1988).

 \bibitem{MacDo} J.R. MacDonald and M. Ausloos, 
  Physica A  {\bf 242}, 150  (1997). 

   
\bibitem{Vicsek} A.-L. Barab\'asi and T. Vicsek,    Phys. Rev. A  {\bf 178}, 2730 (1991).
    
 \bibitem{MA4} M. Ausloos, ÒFinancial Time Series and Statistical MechanicsÓ, in {\it Computational Statistical Physics. From Billiards to Monte Carlo, } K.H. Hoffmann and M. Schreiber, Eds. (Springer-Verlag, Berlin, 2001) pp. 153-168.

 

 \bibitem{0801.2501v1MonaLisa} 
   P. Pedram and G. R. Jafari, 0801.2501v1 
    \bibitem{taylorpollock} R.P. Taylor, A.P. Micolich,  and D. Jonas,
   Nature {\bf 399}, 422  (1999).
  \bibitem{handwriting} N. Vincent, A. Seropian, and G. Stamon, 
       Pattern Recogn. Letters {\bf 26}, 267 (2005).
   \bibitem{Japanesegardens} G. J. van Tonder, 
    Pattern Recogn. Lett. {\bf 28},  728 (2007).
    
 \end{thebibliography}
\end{document}